\documentclass{aastex62}

\received{...}
\revised{...}
\accepted{...}

\submitjournal{ApJ}

\shorttitle{Coronal rain facilitated by magnetic reconnection}
\shortauthors{Li et al.}

\begin{document}

\title{Relation of coronal rain originating from coronal condensations to interchange magnetic reconnection}

\correspondingauthor{Leping Li}
\email{lepingli@nao.cas.cn}

\author[0000-0001-5776-056X]{Leping Li}
\affil{CAS Key Laboratory of Solar Activity, National Astronomical Observatories, Chinese Academy of Sciences, Beijing 100101, People's Republic of China}
\affiliation{Shandong Provincial Key Laboratory of Optical Astronomy and Solar-Terrestrial Environment, and Institute of Space Sciences, Shandong University, Weihai, Shandong 264209, People's Republic of China}
\affiliation{University of Chinese Academy of Sciences, Beijing 100049, People's Republic of China}

\author{Hardi Peter}
\affiliation{Max Planck Institute for Solar System Research, D-37077 G\"{o}ttingen, Germany}

\author[0000-0002-9270-6785]{Lakshmi Pradeep Chitta}
\affiliation{Max Planck Institute for Solar System Research, D-37077 G\"{o}ttingen, Germany}

\author[0000-0001-5705-661X]{Hongqiang Song}
\affiliation{Shandong Provincial Key Laboratory of Optical Astronomy and Solar-Terrestrial Environment, and Institute of Space Sciences, Shandong University, Weihai, Shandong 264209, People's Republic of China}

\begin{abstract}

Using extreme-ultraviolet images, we recently proposed a new and alternative formation mechanism for coronal rain along magnetically open field lines due to interchange magnetic reconnection. In this paper we report coronal rain at chromospheric and transition region temperatures originating from the coronal condensations facilitated by reconnection between open and closed coronal loops. For this, we employ the Interface Region Imaging Spectrograph (IRIS) and the Atmospheric Imaging Assembly (AIA) of the Solar Dynamics Observatory (SDO). Around 2013 October 19, a coronal rain along curved paths was recorded by IRIS over the southeastern solar limb. Related to this, we found reconnection between a system of higher-lying open features and lower-lying closed loops that occurs repeatedly in AIA images. In this process, the higher-lying features form  magnetic dips. In response, two sets of newly reconnected loops appear and retract away from the reconnection region. In the dips, seven events of cooling and condensation of coronal plasma repeatedly occur due to thermal instability over several days, from October 18 to 20. The condensations flow downward to the surface as coronal rain, with a mean interval between condensations of $\sim$6.6 hr. 
In the cases where IRIS data were available we found the condensations to cool all the way down to chromospheric temperatures.
Based on our observations we suggest that some of the coronal rain events observed at chromospheric temperatures could be explained by the new and alternative scenario for the formation of coronal rain, where the condensation is facilitated by interchange reconnection.

\end{abstract}

\keywords{magnetic reconnection --- plasmas
 --- Sun: corona --- Sun: UV radiation --- techniques: spectroscopic}

\section{Introduction} \label{sec:int}

Coronal rain is a well-known phenomenon in the Sun's corona, and a most fascinating features due to its link to coronal heating and coronal magnetic field. It was first discovered in the 1970s \citep{1970PASJ...22..405K, 1972SoPh...25..413L}, and observationally defined as cool and dense blob-like material that forms in a timescale of minutes in the hot corona, and then falls down along curved loop-like paths to the solar surface \citep{2012ApJ...745..152A}. Coronal rain appears frequently in post-flare loops \citep{2003ApJ...586.1417B, 2016ApJ...833..184S} and non-flaring active region closed coronal loops \citep{2001SoPh..198..325S}, with densities ranging in 10$^{10}$-10$^{11}$ cm$^{-3}$ \citep{2015ApJ...806...81A}. It is employed as a tracer of coronal magnetic fields. Using coronal rain observations, a multi-stranded substructure of coronal loops was reported with an average width and length of $\sim$300 km and $\sim$700 km, respectively \citep{2012ApJ...745..152A}. Coronal rain is a recurring (quasi-periodic) phenomenon with the occurrence interval of several hours \citep{2012ApJ...745..152A, 2016ApJ...827...39K, 2018ApJ...853..176A}. The periodicity here may be caused by thermal non-equilibrium \citep[see a review by][]{2020PPCF...62a4016A}.

Coronal rain is mostly seen off-limb as plasma emission in transition region lines, e.g.,\ Si IV, chromospheric lines, e.g., H$\alpha$ and Ca II H, or as absorption in extreme ultraviolet (EUV) images \citep{2001SoPh..198..325S, 2005A&A...443..319D, 2011A&A...532A..96K, 2017A&A...601L...2V}. It is a multithermal structure with temperatures ranging from the transition region ($\sim$10$^{5}$ K) down to chromospheric ($\sim$10$^{4}$ K) temperatures. Coronal rain falls down to the surface with a mean speed of $\sim$70 km s$^{-1}$ in the range of 30-200 km s$^{-1}$ \citep{2001SoPh..198..325S, 2009RAA.....9.1368Z, 2014ApJ...789L..42K, 2016ApJ...833....5S}, and an acceleration of the order of 80$\pm$30 m s$^{-2}$ \citep{2001SoPh..198..325S, 2012ApJ...745..152A}. The acceleration here is considerably smaller than the effective solar gravitational acceleration. One possible explanation for the observed lack in acceleration is that the gas pressure  restructures downstream of the coronal rain \citep{2014ApJ...784...21O}. Moreover, the ponderomotive force from transverse  magnetohydrodynamic (MHD) waves is not expected to play a major role when these waves are present \citep{2017A&A...598A..57V}.  

Two classes of formation mechanisms for coronal rain have been widely investigated. For the non-flaring coronal rain, the heating event, concentrated at (near) the footpoints of active region closed loops, leads to the chromospheric evaporation and direct mass ejections into the loops \citep{2003A&A...411..605M, 2005A&A...436.1067M, 2010ApJ...716..154A, 2015A&A...583A.109L}. The loops thus become hotter and denser quickly, and then rapidly cool due to thermal non-equilibrium, when radiative losses exceed heating input \citep{1953ApJ...117..431P, 1965ApJ...142..531F}. If the heating is sustained and its repetition time is shorter than the radiative cooling time, the loops are brought into a state of thermal non-equilibrium, and undergo thermal non-equilibrium cycles of heating (evaporation) and cooling (condensation) \citep[e.g.,][]{2019A&A...625A.149J}. During the cooling phase, the loops are brought to a state in which thermal instability can be triggered locally \citep{2019SoPh..294..173K, 2020PPCF...62a4016A}. Subsequently, under the effect of gravity, the cool  plasma condensation falls down from the corona to the surface along the loops as coronal rain  \citep{2001SoPh..198..325S, 2012ApJ...745..152A}. In addition, formation of coronal rain along a coronal loop which may be triggered by impulsive heating associated with magnetic reconnection is reported \citep{2019A&A...630A.123K}. For flare-driven coronal rain, the thermal conduction front and beamed nonthermal particles, produced during the flare, heat the chromospheric material quickly, resulting in the chromospheric evaporation and then the greater mass loading of the post-flare loops \citep{2009ApJ...690..347L, 2016NatCo...713798B, 2016ApJ...827...27Z, 2017ApJ...841L...9L, 2017ApJ...836..235L, 2018ApJ...856...34T}. Due to thermal instability, the hot evaporated plasma cools and condenses rapidly, and then falls down along the post-flare loops to the surface as coronal rain \citep{2003ApJ...586.1417B, 2016ApJ...833..184S}. Here, the electron beam heating alone cannot directly produce coronal rain \citep{2020ApJ...890..100R}. These two classes of coronal rains mentioned above have been reported to occur along pre-existing loops that are magnetically closed.

In EUV observations by the Atmospheric Imaging Assembly \citep[AIA;][]{2012SoPh..275...17L} onboard the Solar Dynamics Observatory \citep[SDO;][]{2012SoPh..275....3P} on 2012 January 19, higher-lying curved open loops moved downward and reconnected with the lower-lying closed loops \citep{2018ApJ...864L...4L}. Two sets of newly reconnected loops then formed and retracted away from the reconnection region, indicating the reconfiguration of magnetic field geometry \citep{2010ApJ...723L..28L, 2016ApJ...820L..37C, 2016NatCo...711837X, 2018ApJ...853L..18Y, 2019ApJ...883..104S}. The higher-lying and lower-lying reconnecting loops are rooted in quiet Sun region with weaker magnetic field, and the reconnection lasts for more than 20 hr. Moreover, no associated flare is identified during the reconnection process. Thus, in this event the reconnection rate is quite small. In addition, quasi-periodic fast-mode magnetoacoustic waves originated from the reconnection region, and propagated upward across the higher-lying loops \citep{2018ApJ...868L..33L}. Most of the magnetic energy thus may be converted to wave energy through reconnection. Due to the downward motion a magnetic dip forms in the higher-lying open loops. In the dips, rapid cooling and condensations of coronal plasma take place. Because of the successive reconnection, without support from the underlying loops the coronal condensations fall down to the surface not only along the leg of higher-lying open loops, but also through the reconnection region and down along both legs of newly reconnected closed loops, as coronal rain. Furthermore, over an extended period of time from 2012 January 16 to 20, 15 repeated condensations facilitated by repeated reconnection have been found in the same loop system \citep{2019ApJ...884...34L}. In one of the reconnection events, however, no condensation and subsequent coronal rain were observed. This indicates that not each reconnection event will lead to the formation of condensations. During the reconnection process, cooling of coronal plasma is detected along the higher-lying open loops. In these loops the orientation of the dips is close to the radial direction rather than being horizontal (i.e., parallel to the solar surface), which likely prevented accumulation of material in those structures. If the dip forms in a way that plasma will not accumulate there but still can drain to the surface, it is likely that no condensation can occur. In addition, in the month of 2012 January, 79 other similar events are identified at different positions and times above the limb. This indicates that the condensation facilitated by reconnection between loops (leading to a magnetic dip) is a common phenomenon in the corona. Moreover, so many similar events observed support the solid relation between the reconnection and the condensation for this kind of phenomenon. 

Observations of \citet{2018ApJ...864L...4L, 2018ApJ...868L..33L, 2019ApJ...884...34L} revealed that in the intervals between any two neighboring reconnection and condensation events, the higher-lying curved open loops remain quiescent. They show inclined structures in which no flux rope and no twisted structure is observed. No reconnection between loops, no formation of dips, and no cooling and condensation of coronal plasma is observed. The system of loops appears to be in a state of thermal equilibrium. However, only during the reconnection and condensation events, reconnection results in the magnetic topological changes of loop system. The formation of dips in higher-lying open loops and subsequent newly reconnected loops is observed. During this process, the surrounding plasma is gathered into the dips, leading to the enhancement of plasma density. Triggered by the density enhancement, thermal instability occurs locally, and radiative condensation forms quickly in the dips. Subsequently, the condensation flows to the surface along the leg of higher-lying open loops, and also through the reconnection region along both legs of newly reconnected closed loops, as coronal rain. A new and alternative formation mechanism for coronal rain facilitated by interchange reconnection has been hence suggested \citep{2018ApJ...864L...4L}. Of course, it remains to be seen (from a model point of view), if the reconnection really facilitates the condensation. This model thus needs rigorous testing with numerical simulations, just like the reconnection-condensation model for prominence formation with in situ radiative condensation triggered by reconnection \citep{2015ApJ...806..115K, 2017ApJ...845...12K}. Nevertheless, the reconnection is speeding up the process to actually get a condensation.

Coronal rain events occurring at/near null points and associated dips have been reported previously \citep{2014shin.confE..50L, 2015ApJ...807....7R, 2018cosp...42E3295S}. Recently, coronal rain was observed ubiquitously in active regions within the legs of closed loops inside separatrix domes and null, and at nulls and outer spines. It was suggested that this is caused separately by thermal non-equilibrium or/and via interchange reconnection \citep{2019ApJ...874L..33M}. Different from our results that the coronal rain forms in the dips of higher-lying open loops which attend the interchange reconnection, \citet{2019ApJ...874L..33M} suggest that the coronal rain along the outer spine forms in the newly reconnected open loops. Here, these open loops have a tenuous upper section that forms the outer spine and a hot, dense lower section that forms an arc along the separatrix dome. It is, however, much difficult  to identify the outer spine as the newly reconnected open loops or the reconnecting open loops in \citet{2019ApJ...874L..33M}. Nevertheless, our new formation mechanism for coronal rain along open field lines \citep{2018ApJ...864L...4L} is further supported from the observational and modelling side by \citet{2019ApJ...874L..33M}. All the results point to a new class of coronal rain events due to interchange reconnection under non-flaring conditions \citep{2014shin.confE..50L, 2015ApJ...807....7R, 2018ApJ...864L...4L, 2018ApJ...868L..33L, 2019ApJ...884...34L, 2018cosp...42E3295S, 2019ApJ...874L..33M}, see a review by \citet{2020PPCF...62a4016A}. 

Quite different from coronal rain events along pre-existing magnetically closed loops in non-flaring active regions reported before  \citep{2001SoPh..198..325S, 2010ApJ...716..154A, 2015ApJ...806...81A, 2012ApJ...745..152A}, the new class of quiescent coronal rain previously observed takes place along magnetically open loops \citep{2014shin.confE..50L, 2015ApJ...807....7R, 2018ApJ...864L...4L, 2018ApJ...868L..33L, 2019ApJ...884...34L, 2018cosp...42E3295S, 2019ApJ...874L..33M}. Because of this, the dynamics of these two classes of quiescent coronal rains at coronal heights are quite different. Quiescent coronal rain occurring in the closed loops is mostly observed to fall down toward the surface along one or both legs of the closed loops. Upflow on one loop leg followed by downflow on the other loop leg, and changes of flowing directions along the same loop leg, are also detected \citep[see a review in][]{2020PPCF...62a4016A}. Differing from this picture is quiescent coronal rain occurring in the dips of open loops associated with null point topologies. In this open-loop case, the condensation accumulates, expands, and flows along the supporting field as a prominence, moves together with the evolution of dips, e.g., it sways back and forth parallel to the surface, and also flows in the apparent cross-field direction and across the X-type reconnection point. Aside from these dynamics, downflows, upflows, and counterstreaming of the condensation are also observed \citep[see more details in][]{2018ApJ...864L...4L, 2019ApJ...884...34L}. Another distinguishing feature is that, while  thermal non-equilibrium responsible for  quiescent coronal rain in active region closed loops is promoted by footpoint-concentrated heating \citep{2003A&A...411..605M, 2004A&A...424..289M, 2005A&A...436.1067M, 2010ApJ...716..154A, 2015ApJ...806...81A}, in coronal rain along open loops, it is closely related to the formation of magnetic dips caused by  reconnection \citep{2018ApJ...864L...4L, 2018ApJ...868L..33L, 2019ApJ...884...34L, 2019ApJ...874L..33M} or not associated with reconnection \citep{2006SPD....37.0121L, 2015ASSL..415..205M, 2019ApJ...873...25P}.

Until now, EUV images recorded by SDO/AIA and Solar TErrestrial RElation Observatory (STEREO)/EUV Imager (EUVI) have been  mostly used to investigate the new class of coronal rain \citep{2018ApJ...864L...4L, 2018ApJ...868L..33L, 2019ApJ...884...34L, 2018cosp...42E3295S, 2019ApJ...874L..33M}. \citet{2015ApJ...807....7R} studied a coronal rain event along a dome-shaped structure using the Interface Region Imaging Spectrograph \citep[IRIS;][]{2014SoPh..289.2733D} 1330 \AA~slit-jaw images (SJIs) and Si IV line at 1393.76 \AA, showing the plasma at transition region temperatures. They, however, did not pay attention to the formation of coronal rain, e.g., the cooling and condensation process of coronal plasma. In this study, we investigate the coronal rain, cooling from coronal temperatures all the way down to chromospheric temperatures, facilitated by interchange reconnection. To achieve this, we analyze the evolution of a coronal rain event simultaneously observed by the IRIS and AIA. In short, we can summarize that some of the coronal rain events in transition region and chromospheric lines originate from the condensations facilitated by interchange reconnection between a system of open and closed loops. The observations are shown in Section \ref{sec:obs}. The formation of magnetic dips is discussed in Section \ref{sec:dips}. The results and a summary and discussion are presented respectively in Sections \ref{sec:res} and \ref{sec:sum}. 

\section{Observations}\label{sec:obs}

\begin{figure}[ht!]
\centering
\plotone{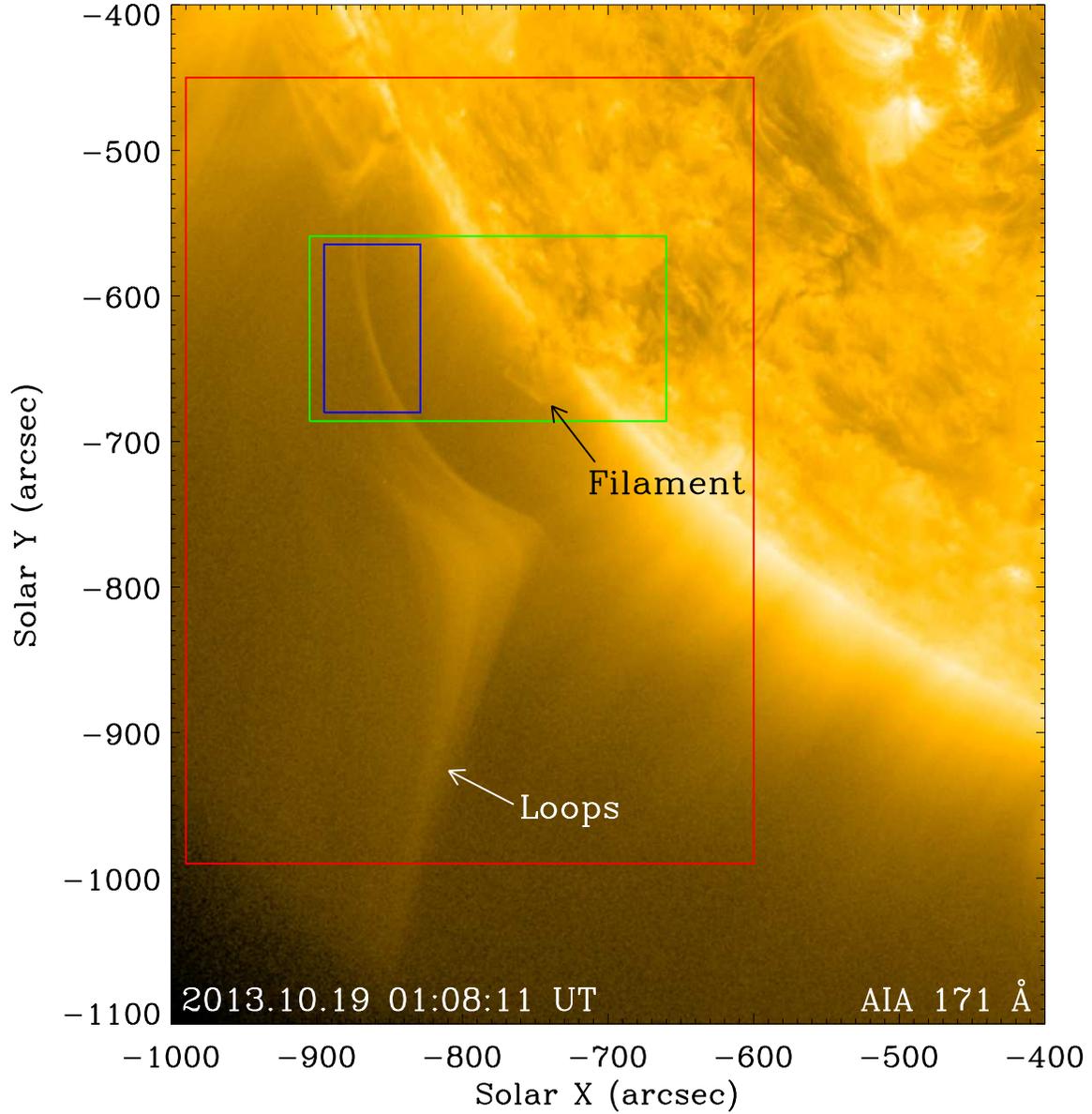}
\caption{Overview of the IRIS and SDO/AIA observations. The background shows an AIA 171 \AA~image from 2013 October 19. The green rectangle shows the field of view (FOV) of the IRIS 1400 \AA~SJIs. The blue and red rectangles separately indicate the FOVs of Figure \ref{f:cr_iris}(a) and Figures \ref{f:mr}(a)-(b), \ref{f:cc}, \ref{f:cr_aia}(a), and \ref{f:composite}(a). The filament is also marked as it was the primary observing target of IRIS at that time. See Section \ref{sec:obs} for details.
\label{f:general}}
\end{figure}

The IRIS observatory provides simultaneous images and spectra of the temperature minimum region, chromosphere, transition region, and corona \citep{2014SoPh..289.2733D}. From 23:59 UT on 2013 October 18 to 03:20 UT on October 19, IRIS observed a filament at the southeastern solar limb, see Figure \ref{f:general}. The field of view (FOV) of IRIS SJIs is indicated by the green rectangle in Figure \ref{f:general}. Six raster scans with the slit spectrograph are acquired, in which the large coarse 64-step raster mode is employed with slit width of 0.33\arcsec, step size of 2\arcsec, and exposure time of 30 s for each raster step. For the SJIs, only the 1400 \AA~wavelength band is recorded. This covers the transition region Si IV lines at 1393.76 \AA~and 1402.77 \AA. We use IRIS level 2 data (available at https://iris.lmsal.com) which are already corrected for flat field, geometric distortions, and dark current. The general information on the IRIS observations is listed in Table \ref{tab:iris}.  The coronal rain we concentrate on in this study is located to the east of the filament, and its FOV is enclosed by the blue rectangle in Figure \ref{f:general}.

To place the IRIS data in the context of the structure and evolution of the corona, we use EUV images from the AIA; see Figure \ref{f:general}. The time cadence and spatial sampling of the AIA EUV images are 12 s and 0.6\arcsec/pixel, respectively. We scale the AIA images to match the IRIS SJIs, and co-align them using several characteristic features, such as the filament and plage patterns. In this study, AIA 171 \AA, 131 \AA, and 304 \AA~images are employed, which show plasma with different characteristic temperatures, i.e., 171 \AA~peaks at $\sim$0.9 MK (Fe IX), 131 \AA~peaks at $\sim$0.6 MK (Fe VIII) and $\sim$10 MK (Fe XXI), and 304 \AA~peaks at $\sim$0.05 MK (He II). Here, the He II, dominating the 304 \AA~channel, and Si IV lines form at similar temperatures.

\section{The formation of magnetic dips}\label{sec:dips}

As we have shown in earlier papers \citep{2018ApJ...864L...4L, 2018ApJ...868L..33L, 2019ApJ...884...34L}, first we see the formation of a dip in the 171 \AA~channel with the signatures typical for reconnection. This is signified by the inward motion to the reconnection site of the loops that form a dip and the outward motion of newly-formed loops away from the reconnection site. Only after this process started, our observations display signatures in the 131 \AA~and 304 \AA~channels showing the cooler condensed plasma. We thus consider this as sufficient observational evidence that the reconnection first forms the dip. Only then (initially hot) plasma accumulates in the dip and subsequently condenses and finally rains down along the loops. In principle, the magnetic dips could also be due to the flux ropes or other twisted structures that lead to the condensations \citep[as in some prominence formation models, e.g.,][]{1983SoPh...88..219P}. However, in our observations, we do not see any obvious evidence for the presence of flux ropes or twisted structures. The higher-lying open loops simply show inclined structures before the dip formation \citep{2018ApJ...864L...4L, 2018ApJ...868L..33L, 2019ApJ...884...34L}. It is thus unlikely that the magnetic dips are related to flux ropes or other twisted structures.

In principle, one could also envision a scenario where mass loading on a field line would lead to the formation of a dip in the field line \citep[like in some prominence formation models; e.g.,][]{2001ApJ...553L..85K}. The condensed material would then push the field line down towards a lower-lying oppositely directed field line and drive forced reconnection \citep{2005GeoRL..32.6105B}. However, to form such a dip in a field line, prominence formation models need to start with an almost horizontal field line to prevent the plasma from quickly draining back to the surface without sufficient mass loading to form a dip. In the cases we investigated, the off-limb observations show that the higher-lying open loops, on which the dip forms, are far from being horizontal \citep{2018ApJ...864L...4L, 2018ApJ...868L..33L, 2019ApJ...884...34L}. It is thus unlikely that the dip formation is initiated by the condensation. Furthermore, if the condensation would form first (independent of the reconnection) and then form the magnetic dip, we would see the 304 \AA~emission before the dip formation, or at least together with the formation of the dip. However, this is not the case and first the dip forms in the 171 \AA~channel and later the 304 \AA~emission appears. 

Based on this, we conclude that reconnection between loops forms the magnetic dips in the higher-lying open loops, and facilitates the condensations of coronal plasma, that then fall down to the surface as coronal rain. Still, it is likely that once condensations form, their presence will facilitate a speed-up of the reconnection.

\section{Results}\label{sec:res}

\subsection{Coronal rain observed by IRIS}\label{sec:cr_iris}

The coronal rain appeared in the IRIS 1400 \AA~SJIs starting from $\sim$00:05 UT on 2013 October 19, and moved toward the north along curved loop-like paths; see the online animated version of Figure \ref{f:cr_iris}. Fine structures of the coronal rain are identified with width and length of $\sim$700 km and $\sim$90 Mm, respectively; see Figure \ref{f:cr_iris}(a). Along the moving direction AB in the green box in Figure \ref{f:cr_iris}(a), a time-slice of IRIS 1400 \AA~SJIs is made and displayed in Figure \ref{f:cr_iris}(b). Multiple flows of the coronal rain are clearly detected with a mean speed of $\sim$40 km s$^{-1}$; see the cyan dotted line in  Figure \ref{f:cr_iris}(b).

Due to the small FOV, IRIS only scans part of the coronal rain in the south, enclosed by the pink rectangle in Figure \ref{f:cr_iris}(a), and only during the second to the fifth raster scans; see the online animated version of Figure \ref{f:cr_iris}. For the spectroscopic analysis, we concentrate on the transition region lines of Si IV at 1394 \AA~($\sim$0.06 MK) and chromospheric lines of Mg II k at 2796 \AA~($\sim$0.01 MK) \citep{2014SoPh..289.2733D}. The alignment between these two lines is achieved using the fiducial marks on the slit. For the coronal rain, emission profiles of both the Si IV and Mg II k lines are obtained; see the diamond signs in Figures \ref{f:lp}(a)-(b). Both profiles are well approximated by single-Gaussian fits with constant background for the continuums (see the red lines in Figures \ref{f:lp}(a)-(b)) to build maps of intensity, Doppler shift, and line width. The line ratios of Si IV 1394 \AA~and 1403 \AA~are measured to be 1.6-2.2 with a mean value of 1.9, and those of Mg II 2796 \AA~and 2803 \AA~are measured to be 1.5-2.0 with a mean value of 1.7. These results indicate that the coronal rain is optically thin.

For the Doppler maps, wavelength calibrations are required for both of the two lines; see Figures \ref{f:gf}(b1)-(b2). These are also provided with the IRIS level 2 data, but here we check and reiterate the calibrations. The raster maps covered a portion of the quiet Sun near the southeast limb. Spectral profiles from this quiet Sun region from neutral and singly ionized atomic species, e.g., S I, Fe II, and Ni II lines for Si IV (1394 \AA) line and Mn I, Fe I, and Ni I lines for Mg II k (2796 \AA) line, are used for the wavelength calibrations \citep{2014A&A...570A..93L, 2014Sci...346C.315P, 2017RAA....17..110T, 2019A&A...626A..98L}. Here, the uncertainty in the wavelength calibration in Doppler units is expected to be about 1 km s$^{-1}$. Information of the two target lines and their associated calibration lines is listed in Table \ref{tab:lines}. After the wavelength calibrations, the Si IV and Mg II k lines have redshifts of 2.3 km s$^{-1}$ and 3.1 km s$^{-1}$, respectively, in the quiet Sun region. Here, the Doppler shift of Mg II k line is obtained using the central absorption at k3, and unrelated to the spectroscopic results of coronal rain.

Similar spectroscopic results of coronal rain are obtained for the four middle raster scans of IRIS, i.e., from scans number two to five. Among them we take the results for the third raster scan for example, and display them in Figure \ref{f:gf}. The left and right columns of Figure \ref{f:gf} show the maps of intensity, Doppler shift, and line width of the Si IV and Mg II k lines, respectively. Here, the instrumental line width of about 26 m\AA~is subtracted from the Gaussian line width \citep{2014SoPh..289.2733D}. In these maps, the line fit parameters outside the locations of the coronal rain are not shown as they are not well-fitted because of low count rates. Therefore, we only concentrate on the coronal rain region enclosed by the blue dash-dotted lines in Figure \ref{f:gf}. Same as in the IRIS 1400 \AA~SJIs (see Figure \ref{f:cr_iris}(a)), two substructures of the coronal rain are identified, marked separately by the blue square and diamond signs in Figure \ref{f:gf}. This demonstrates that in our observations in the coronal rain, the 1400 \AA~channel is indeed dominated by the transition region Si IV line. Comparing the Si IV and Mg II k intensity maps, we find that the coronal rain seen in the Mg II k line is located mostly at the same locations as the Si IV coronal rain; see Figures \ref{f:gf}(a1)-(a2). At the north (south) coronal rain, we measure the intensity, Doppler shift, and line width, and obtain the mean values of 46 (18) DN pixel$^{-1}$, 7.5 (5) km s$^{-1}$, and 8.6 (7.3) km s$^{-1}$ for the Si IV line, and 218 (122) DN pixel$^{-1}$, 8.7 (5.9) km s$^{-1}$, and 7.9 (8.3) km s$^{-1}$ for the Mg II k line. Here, the Doppler shift with a positive value shows the redshift. Moreover, the intensity of the Mg II k line is much larger than that of the Si IV line; see Figures \ref{f:gf}(a1)-(a2). The Doppler shift and line width are, nevertheless, similar for both of the two lines; see Figures \ref{f:gf}(b1)-(b2) and (c1)-(c2).

\subsection{Coronal condensations facilitated by magnetic reconnection observed by AIA}\label{sec:cr_sdo}

\begin{figure}[ht!]
\centering
\includegraphics[width=\textwidth]{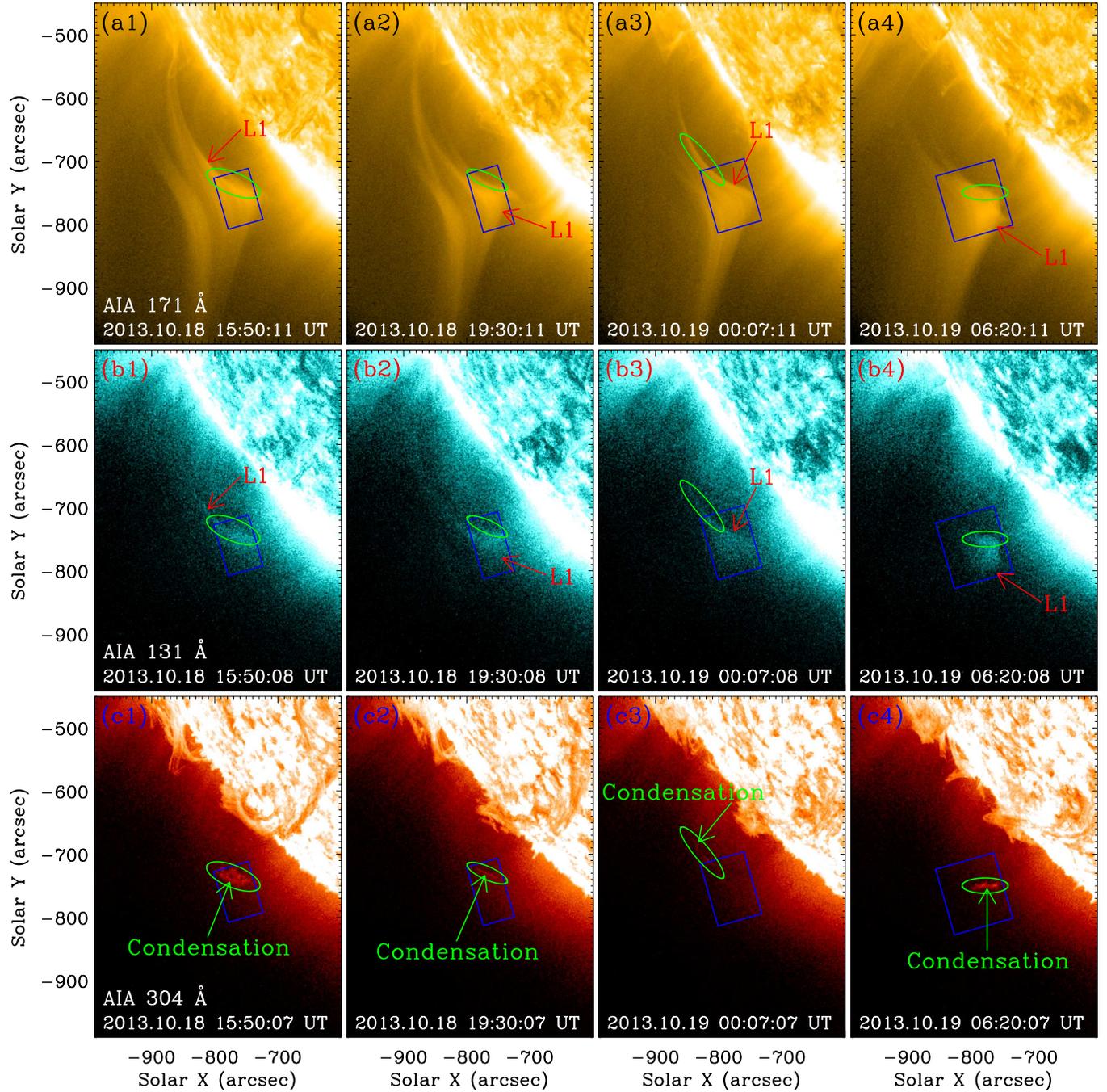}
\caption{Cooling and condensations of coronal plasma observed by AIA. (a1)-(a4) AIA 171 \AA, (b1)-(b4) 131 \AA, and (c1)-(c4) 304 \AA~images. The green ellipses in (a1)-(c1), (a2)-(c2), (a3)-(c3), and (a4)-(c4) separately enclose the condensations in (c1), (c2), (c3), and (c4). The blue rectangles in (a1)-(c1), (a2)-(c2), (a3)-(c3), and (a4)-(c4) show the regions for the light curves of the AIA 171 \AA, 131 \AA, and 304 \AA~channels as displayed in Figures \ref{f:lc}(a), (b), (c), and (d) by red, green, and blue lines, respectively. The FOV is denoted by the red rectangle in Figure \ref{f:general}. An animation of the unannotated SDO/AIA images is available. It covers 27 hr starting from 10:00 UT on 2013 October 18, with a cadence of 1 min. See Section \ref{sec:cr_sdo} for details. (An animation of this figure is available.)
\label{f:cc}}
\end{figure}

To the east of the filament scanned by the IRIS rasters, a system of higher-lying curved open coronal loops are observed in AIA 171 \AA~images; see Figure \ref{f:general}. We investigate the evolution of loops over an extended period of time from 2013 October 18 to 20. Similar to \citet{2018ApJ...864L...4L, 2018ApJ...868L..33L, 2019ApJ...884...34L}, the higher-lying open loops, denoted by L1, show inclined structures in which no flux rope and no twisted structure is observed, see the online animated version of Figure \ref{f:cc}. They move repeatedly to the west toward the surface; see Figure \ref{f:mr}(a). Magnetic dips then form in loops L1; see Figure \ref{f:mr}(a). The downward-moving loops L1 meet the lower-lying closed loops, marked as L2, forming an X-type configuration; see Figure \ref{f:mr}(b). At the interface, reconnection between loops L1 and L2 takes place; see Figure \ref{f:mr}(b). As a consequence, two sets of newly reconnected loops, denoted by L3 and L4, form; see Figure \ref{f:mr}(b). No associated flare is detected during the reconnection process. Moreover, some of the loops L1 move upward away from the reconnection region between loops L1 and L2, see the online animated version of Figure \ref{f:cc}. These loops may be freed from the heavy material and recede upward due to the magnetic tension, when the reconnection occurs and the material is allowed to cross the null point. Along the CD direction in the blue rectangle in Figure \ref{f:mr}(a), a time-slice of AIA 171 \AA~images is made and displayed in Figure \ref{f:mr}(c). Multiple motions of loops L1 are clearly detected, with speeds in the range of 4-5 km s$^{-1}$; see the red dotted lines in Figure \ref{f:mr}(c). Quasi-periodic disturbances propagating upward across the dip of loops L1 are detected, see Figure \ref{f:mr}(c). They are likely to represent the quasi-periodic fast propagating magnetoacoustic waves originating from the reconnection region between loops L1 and L2 \citep{2018ApJ...868L..33L}. Along the EF direction in the green rectangle in Figure \ref{f:mr}(b), another time-slice of AIA 171 \AA~images is obtained and shown in Figure \ref{f:mr}(d). The newly reconnected loops L3 and L4 retract away from the reconnection region after the appearance with speeds in the range of 7-15 km s$^{-1}$; see the green dotted and dashed lines in Figure \ref{f:mr}(d).

Along with the motions of loops L1, bright emission appears in AIA 131 \AA~images and moves together with the 171 \AA~loops; see Figures \ref{f:cc}(b1)-(b4). Same as in \citet{2018ApJ...864L...4L, 2018ApJ...868L..33L, 2019ApJ...884...34L}, no associated bright emission is observed in other AIA wavelength channels that show plasma at higher characteristic temperatures, e.g., 193 \AA, 211 \AA, 335 \AA, and 94 \AA. Therefore, the emission we see here in the 131\,{\AA} channel shows plasma with the lower ($\sim$0.6 MK) rather than the higher ($\sim$10 MK) characteristic temperature of the 131 \AA~channel. In particular, the plasma we see in the 131\,{\AA} channel will be cooler than what we see in AIA 171 \AA~($\sim$0.9 MK) loops.

Bright emission appears repeatedly in AIA 304 \AA~images, which show plasma at the characteristic temperature of $\sim$0.05 MK. As listed in Table \ref{tab:cc}, we see seven condensations of coronal plasma; see also the online animated version of Figure \ref{f:cc}. Among them, the first four condensations are displayed separately in Figures \ref{f:cc}(c1)-(c4). We overlay these condensations on the AIA 171 \AA~and 131 \AA~images, and find that all of them take place at the north edge of dips near the endpoints of loops L1; see the green ellipses in Figure \ref{f:cc}. 

\begin{figure}[ht!]
\centering
\includegraphics[width=0.5\textwidth]{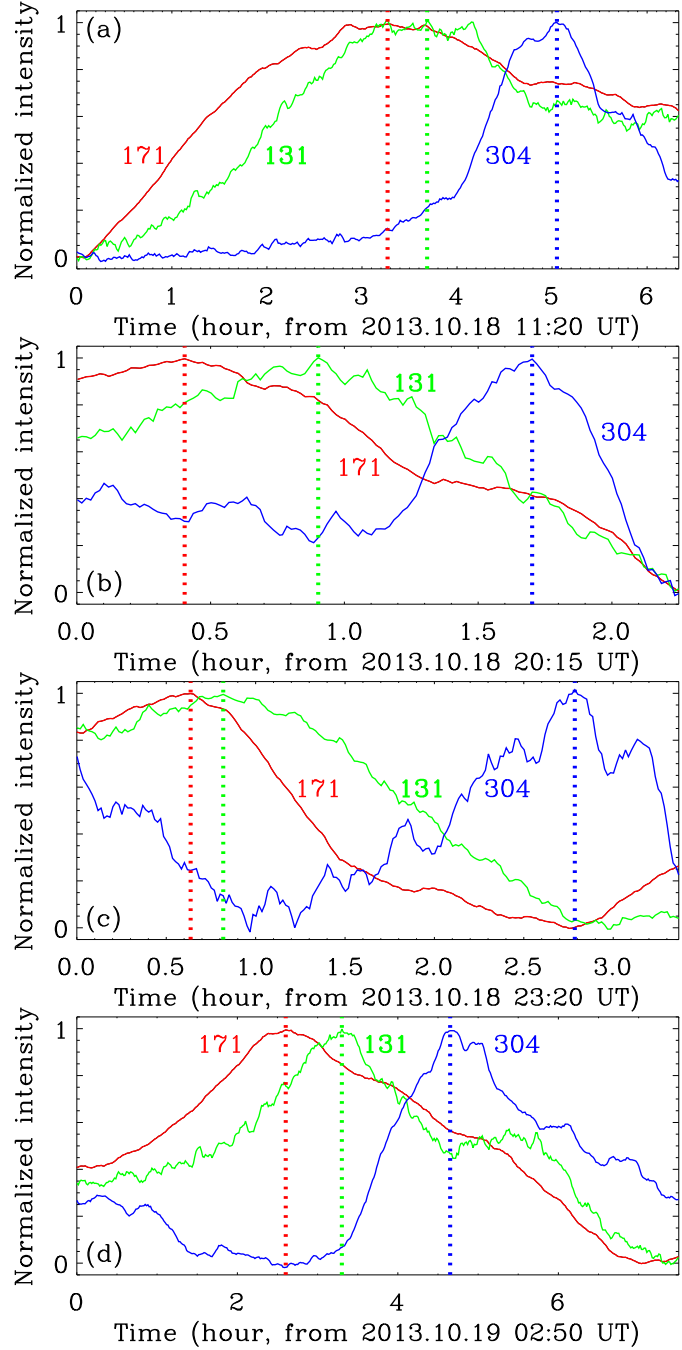}
\caption{Temporal evolution of coronal condensations observed by AIA. (a)-(d) Light curves of the AIA 171 \AA~(red lines), 131 \AA~(green lines), and 304 \AA~(blue lines) channels in the blue rectangles in Figures \ref{f:cc}(a1)-(c1) (a), (a2)-(c2) (b), (a3)-(c3) (c), and (a4)-(c4) (d), respectively. The red, green, and blue vertical dotted lines separately mark the peaks of the AIA 171 \AA, 131 \AA, and 304 \AA~light curves. See Section \ref{sec:cr_sdo} for details.
\label{f:lc}}
\end{figure}

In the blue rectangles in Figure \ref{f:cc}, light curves of the AIA 171 \AA, 131 \AA, and 304 \AA~channels for these four events are calculated and displayed in Figures \ref{f:lc}(a)-(d) as the red, green, and blue lines, respectively. For each event, all the three light curves increase first, reach the peaks, and then decrease. They, however, peak at different times; see the red, green, and blue vertical dotted lines in Figure \ref{f:lc}. The ordering of the peaks is always from 171 {\AA} to 131 {\AA} to 304 {\AA}, i.e., in the order of decreasing formation temperature of the respective channels. Therefore, we can clearly identify these cases as cooling and condensation processes of coronal plasma in the magnetic dips. 

Due to the successive reconnection process between loops L1 and L2, the condensations lost support by the underlying loops L1. Consequently they fall to the surface not only along the leg of loops L1, but also through the reconnection region along both legs of newly reconnected closed loops L4, as coronal rain; see Figure \ref{f:cr_aia}(a). Along the moving direction GH in the blue rectangle in Figure \ref{f:cr_aia}(a) we calculate a time-slice of AIA 304 \AA~images and show this in Figure \ref{f:cr_aia}(b). Multiple downflows along the south leg of loops L4 are obvious, with speeds of $\sim$70 km s$^{-1}$; see the green dotted line in Figure \ref{f:cr_aia}(b). Along another moving direction IJ in the green box in Figure \ref{f:cr_aia}(a), we make another time-slice of AIA 304 \AA~images and display it in Figure \ref{f:cr_aia}(c). Condensations fall repeatedly along the north leg of loops L4 and the leg of loops L1, with speeds of $\sim$60-80 km s$^{-1}$, see the blue and green dotted lines in Figure \ref{f:cr_aia}(c). As a consequence, the condensation near the reconnection region disappears. 

Parameters, including the start time, the end time, the cooling times from 171 \AA~to 131 \AA~and then to 304 \AA, the appearance time, the peak time, the disappearance time, and the lifetime, of the seven condensations are listed in Table \ref{tab:cc}. The coronal plasma in the dips cools down from $\sim$0.9 MK, the characteristic temperature of AIA 171 \AA~channel, to $\sim$0.6 MK, the characteristic temperature of AIA 131 \AA~channel, in 11-51 minutes with a mean value of 33 minutes, and then to $\sim$0.05 MK, the characteristic temperature of AIA 304 \AA~channel, in another 48-174 minutes with an average value of 110 minutes. The lifetimes of condensations are distributed over a range from 4.5 to 8.1 hr, with a mean value of 6.2 hr. Using the duration between two neighboring peak times of the AIA 304 \AA~light curves, we calculate the occurrence interval of condensations, and thus of coronal rain, and obtain an average value of 6.6 hr in the range of 4.2-10.4 hr.

\subsection{Relation between coronal rain observed by IRIS and AIA}\label{sec:relation}

The FOV of coronal rain observed by IRIS covers part of the north leg of loops L1, and thus also L4; see the blue rectangle in Figure \ref{f:general}. Moreover, the observing time of IRIS is in the time range of the third condensation event recorded by AIA; see Tables \ref{tab:iris} and \ref{tab:cc}. A composite of IRIS 1400 \AA~SJIs (red) and AIA 171 \AA~(green) and 304 \AA~(blue) images at 01:07 UT on October 19 is made and displayed in Figure \ref{f:composite}(a). It shows that the coronal rain in IRIS is co-temporal and co-spatial with the flow of condensations, and thus the coronal rain, in AIA 304 \AA~images, along the leg of loops L1 and/or the north leg of loops L4; see the online animated version of Figure \ref{f:composite}. Due to the different spatial resolutions of IRIS and AIA, and the limited overlapping time period, it is not clear to what extent the coronal rain in IRIS 1400 \AA~SJIs is co-spatial with that in AIA 304 \AA~images. Furthermore, we overlay the time-slice of IRIS 1400 \AA~SJIs shown in Figure \ref{f:cr_iris}(b) on that of AIA 304 \AA~images in Figure \ref{f:cr_aia}(c) as a small panel. Same flows are detected in both time slices, see Figure \ref{f:cr_aia}(c). For example, comparing the cyan dotted lines in Figures \ref{f:cr_iris}(b) and \ref{f:cr_aia}(c), we find a similar evolution of coronal rain in both the IRIS 1400\,{\AA} SJIs and AIA 304\,{\AA} images. After moving out of the IRIS FOV, the coronal rain continues accelerating as before from $\sim$40 km s$^{-1}$ to $\sim$60 km s$^{-1}$ toward the surface, with an acceleration of $\sim$16 m s$^{-2}$; see the blue dotted line in Figure \ref{f:cr_aia}(c).

In the black rectangle in Figure \ref{f:composite}(a), we calculate the light curves of coronal rain in the AIA 304 \AA~and IRIS 1400 \AA~channels and display them in Figure \ref{f:composite}(b) with blue and red lines. Due to the IRIS observing mode (see Section \ref{sec:obs}), we can investigate the coronal rain in the IRIS SJIs only during the times of the six raster scans; see the red lines in Figure \ref{f:composite}(b). Moreover, for the IRIS 1400 \AA~light curve, the FOV changes when the right or left edges of IRIS 1400 \AA~SJIs pass through the black rectangle. To avoid this problem, we calculate the mean, rather than the total, intensity of bright coronal rain alone, without the dark background, in the FOV for both light curves. If the intensity of coronal rain does not vary strongly at different positions in the black rectangle when the FOV changes, the IRIS 1400 \AA~light curve then depends weakly on the changing FOV. In spite of these limitations, a similar temporal evolution of the two light curves is clearly found, which reflects that the 304 \AA~(He II) and 1400 \AA~(Si IV) channels form at similar temperatures. Our results suggest that the coronal rain here observed by IRIS originates from the condensations facilitated by interchange reconnection between loops L1 and L2 observed by AIA.

\section{Summary and discussion}\label{sec:sum}

Using the SDO/AIA and IRIS images and spectra we study coronal rain that is facilitated by reconnection events. Here we concentrate on repetitive rain at the southeastern solar limb around 2013 October 19. In the IRIS 1400 \AA~SJIs the coronal rain moves down along curved loop-like paths with speeds of $\sim$40 km s$^{-1}$. Only a small part of coronal rain in the south is scanned by the IRIS rasters, while the AIA data cover the full spatial and temporal extent. We analysed emission line profiles of both Si IV and Mg II k at the locations of the coronal rain. To the line profiles we fitted single-Gaussians with a constant continuum to build maps in intensity, Doppler shift, and line width. The rain as seen in Si IV and in Mg II appears at the same position, at least in almost all cases. This, of course, needs to be tested by statistical studies in the future. Moreover, in the coronal rain we find similar Doppler shifts and line widths for both of the two lines. 

\begin{figure}[ht!]
\centering
\includegraphics[width=0.9\textwidth]{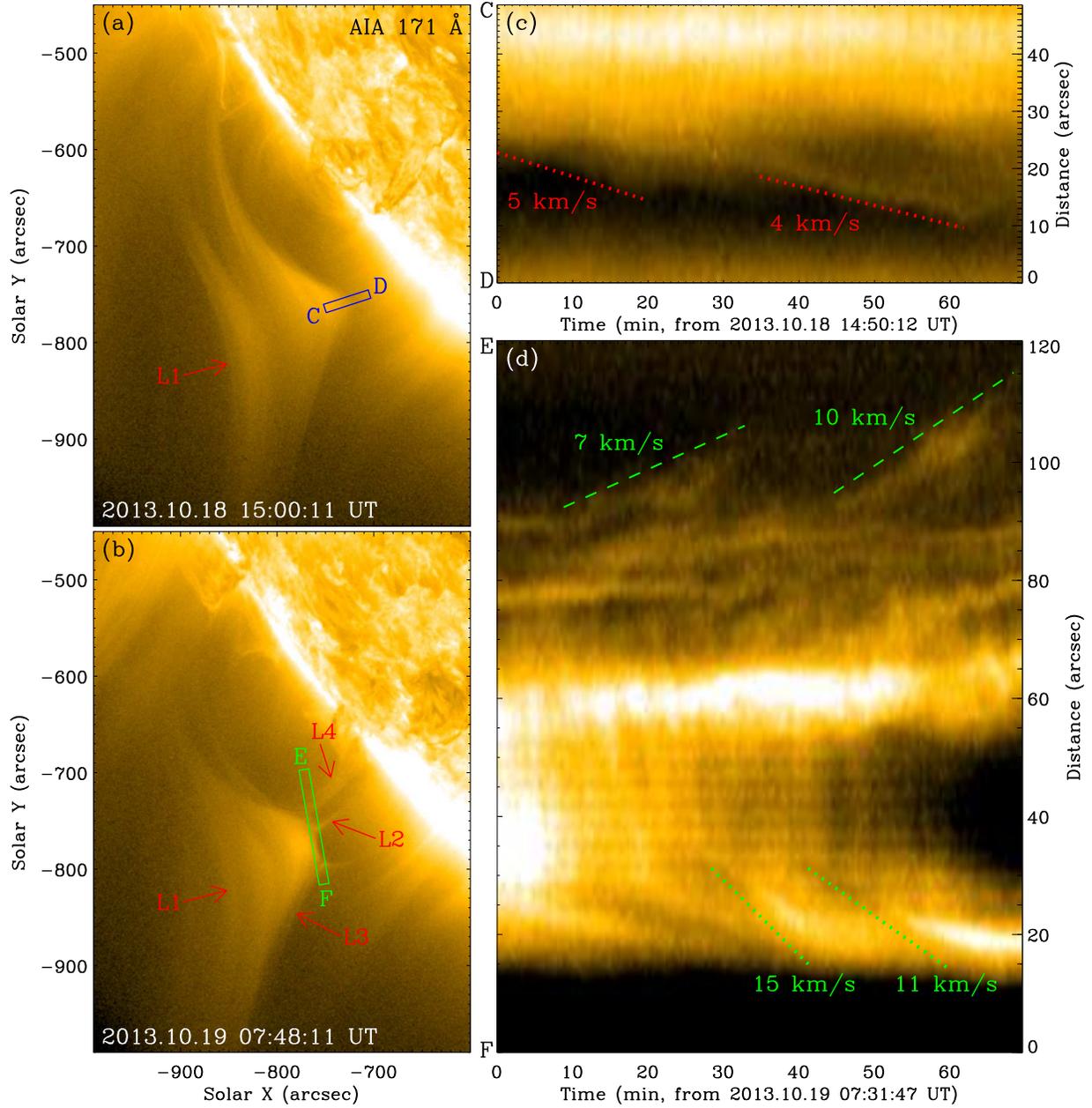}
\caption{Temporal evolution of magnetic reconnection between coronal loops observed by SDO/AIA. (a)-(b) AIA 171 \AA~images and (c)-(d) time-slices of AIA 171 \AA~images along the CD and EF directions in the blue and green rectangles in (a) and (b), respectively. The red dotted lines in (c) indicate the downward motions of loops L1. The green dotted and dashed lines in (d) separately indicate the retraction of the newly reconnected loops L3 and L4 away from the reconnection region between loops L1 and L2. The speeds of these loops are denoted by the numbers in (c) and (d). The FOV of (a)-(b) is indicated by the red rectangle in Figure \ref{f:general}. See Section \ref{sec:cr_sdo} for details.
\label{f:mr}}
\end{figure}

In the corona surrounding the coronal rain, a set of higher-lying curved open loops (L1) are observed in AIA 171 \AA~images (see Figure \ref{f:mr} for the nomenclature of the loops). They move downward towards the surface, and meet lower-lying closed loops (L2). The EUV observations in the 171\,{\AA} channel suggest that between them interchange reconnection repeatedly takes place. Consequently, magnetic dips form in the higher-lying loops (L1). In the same process, two sets of newly reconnected loops (L3 and L4) appear and retract away from the reconnection region. In the dips of the loops L1, bright emission appears subsequentially in AIA 131 \AA~and 304 \AA~images indicating a cooling of the coronal plasma from almost 1\,MK as seen at 171\,{\AA} to below 0.1\,MK in 304 {\AA}.  Thus condensations of cold plasma form. Without support from the underlying loops L1, the condensations flow to the surface along both legs of the newly formed closed loops L4, and also the leg of higher-lying open loops L1, as coronal rain. In the AIA images, the coronal rain recurs for seven times with a mean occurrence interval of $\sim$6.6 hr spanning from 4.2-10.4 hr. During the third of these seven cases, the coronal rain is caught also by the raster scan of IRIS. Here the rain as seen in Si IV and in Mg II is co-spatial and co-temporal with the coronal rain as seen in AIA. As we mentioned in Section \ref{sec:relation}, to what extent the rain in Si IV and Mg II k lines is co-spatial with that in AIA 304 \AA~images is not clear. Statistical studies are thus needed in the future.

\begin{figure}[ht!]
\plotone{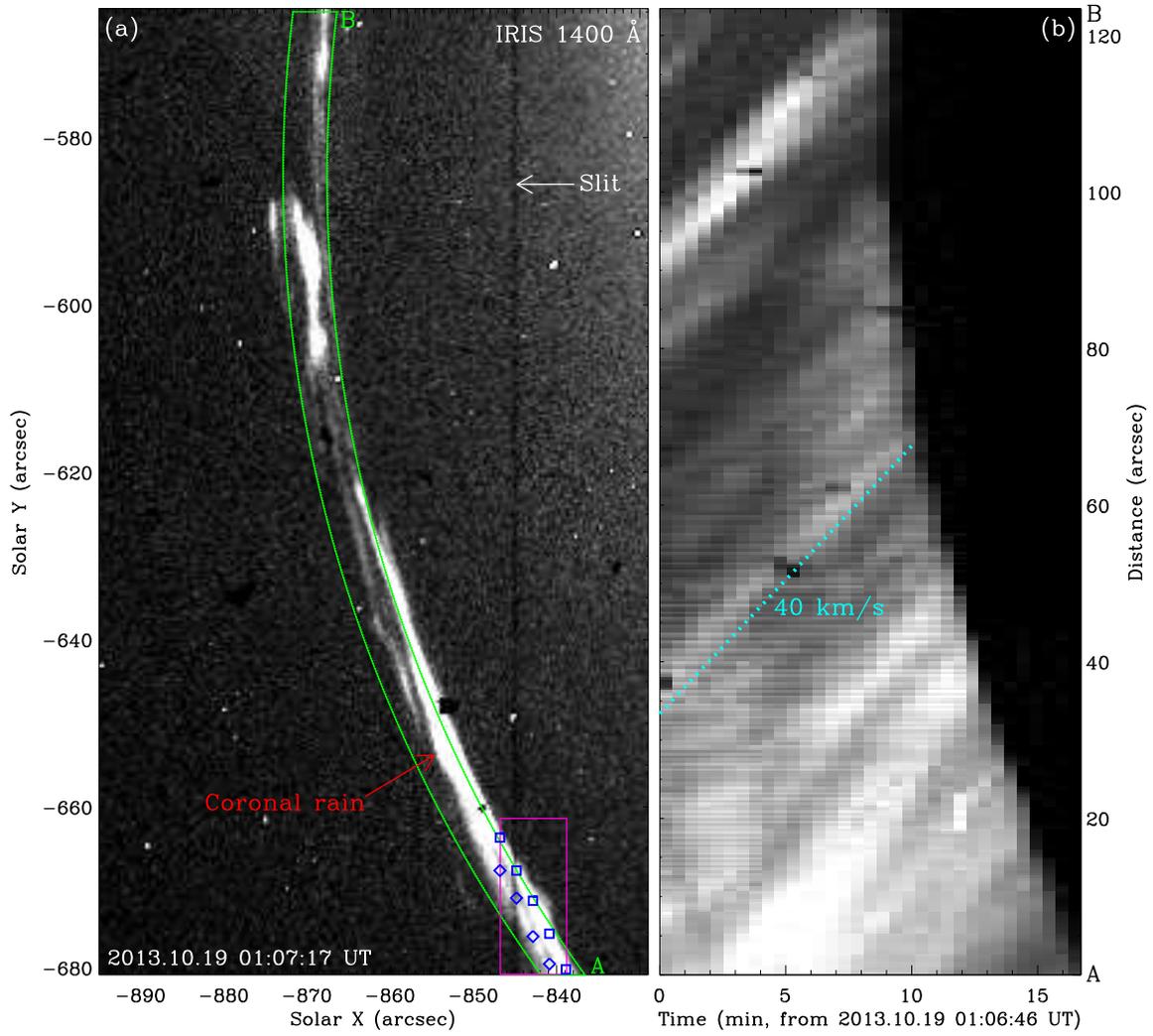}
\centering
\caption{Coronal rain observed by IRIS. (a) An IRIS SJI at 1400 {\AA} (North is top) and (b) the time-slice of IRIS 1400 \AA~SJIs along the AB direction in the green box in (a). The pink rectangle in (a) denotes the FOV of Figure \ref{f:gf}. The blue squares and diamonds in (a) separately mark the positions of northern and southern parts of the coronal rain scanned by the IRIS slit. The cyan dotted line in (b) indicates a speed of 40 km s$^{-1}$ that corresponds to the motion of the coronal rain. The FOV of (a) is denoted by the blue rectangle in Figure \ref{f:general}. An animation of the unannotated IRIS 1400 \AA~SJIs (panel (a)) is available online. It covers 3 hr for the six raster scans starting at 23:59:51 UT on 2013 October 18, and the video cadence is 31.4 s for each raster scan. See Section \ref{sec:cr_iris} for details. (An animation of this figure is available.)
\label{f:cr_iris}}
\end{figure}

\begin{figure}[ht!]
\centering
\includegraphics[width=0.9\textwidth]{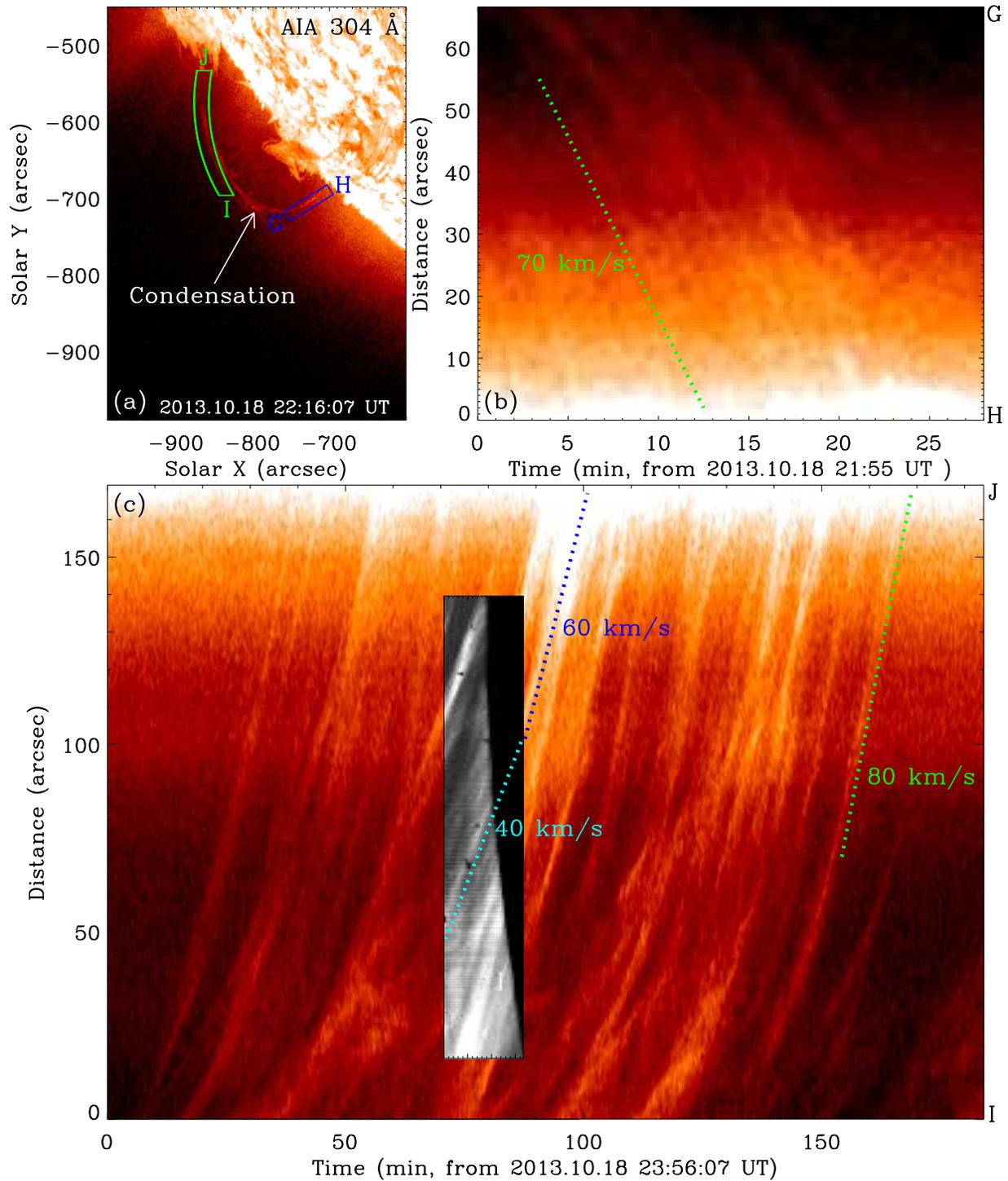}
\caption{Coronal rain observed by AIA. (a) An AIA 304 \AA~image and (b)-(c) time-slices of AIA 304 \AA~images along the GH and IJ directions in the blue and green boxes in (a). The green, cyan, and blue dotted lines in (b) and (c) outline the downward motions of coronal rain. The small panel in (c) shows the time slice of IRIS 1400 \AA~SJIs in Figure \ref{f:cr_iris}(b). The FOV of (a) is denoted by the red rectangle in Figure \ref{f:general}. See Sections \ref{sec:cr_sdo} and \ref{sec:relation} for details.
\label{f:cr_aia}}
\end{figure}

\begin{figure}[ht!]
\centering
\includegraphics[width=0.66\textwidth]{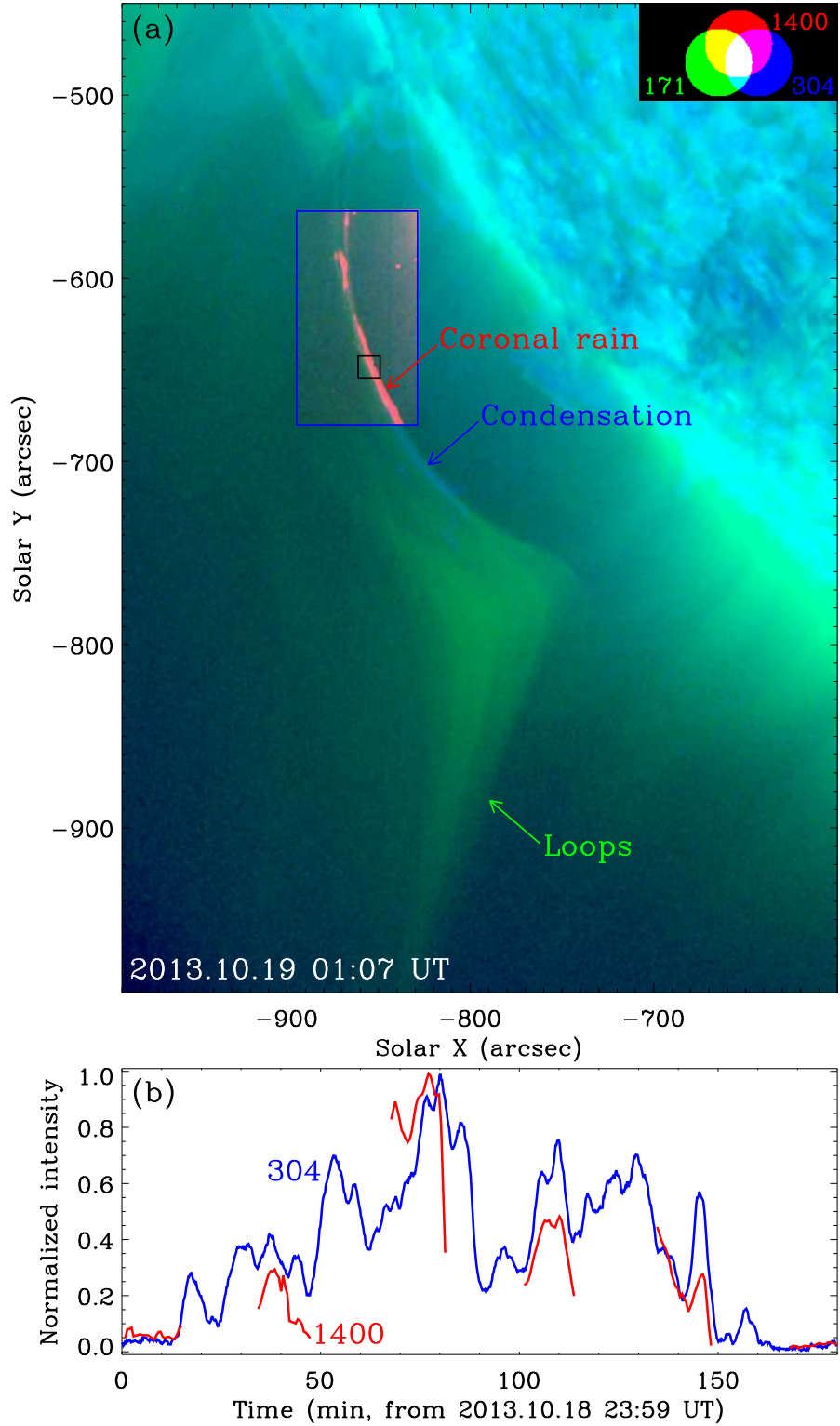}
\caption{Spatial and temporal relation between the coronal rain observed by IRIS and AIA. (a) A composite of AIA 171 \AA~(green), 304 \AA~(blue), and IRIS 1400 \AA~slit-jaw (red) images. (b) Light curves of the AIA 304 \AA~(blue) and IRIS 1400 \AA~(red) channels in the black rectangle in (a). Same as in Figure \ref{f:general}, the blue rectangle in (a) indicates the FOV of Figure \ref{f:cr_iris}(a). The FOV of (a) is denoted by the red rectangle in Figure \ref{f:general}. An animation of the unannotated composite images (panel (a)) is available. It covers 3 hr for the six IRIS raster scans starting at 23:59:51 UT on 2013 October 18, and the video cadence is 31.4 s for each raster scan. See Section \ref{sec:relation} for details. (An animation of this figure is available.)
\label{f:composite}}
\end{figure}

The coronal rain we report here is not related to flaring activity. It clearly appears in the IRIS 1400 \AA~SJIs (see Figure \ref{f:cr_iris}a). This indicates that the coronal rain contains plasma at transition region temperature of below 0.1 MK (Si IV). We find fine structures of the coronal rain (see Section \ref{sec:cr_iris}) supporting the multi-stranded loop scenario \citep{2012ApJ...745..152A}. For the width of the coronal rain we measured a value of $\sim$700 km (see Section \ref{sec:cr_iris}), identical to the coronal rain widths in \citet{2015ApJ...806...81A}. The coronal rain moves along curved loop-like paths with a mean speed of $\sim$40 km s$^{-1}$. As measured along the higher part of the loops, the speed here is small, but similar to those reported previously \citep{2001SoPh..198..325S, 2009RAA.....9.1368Z, 2012ApJ...745..152A, 2015ApJ...806...81A}. The coronal rain continues accelerating after it moves out of the IRIS FOV, see Figure \ref{f:cr_aia}(c). 

\begin{figure}[ht!]
\plotone{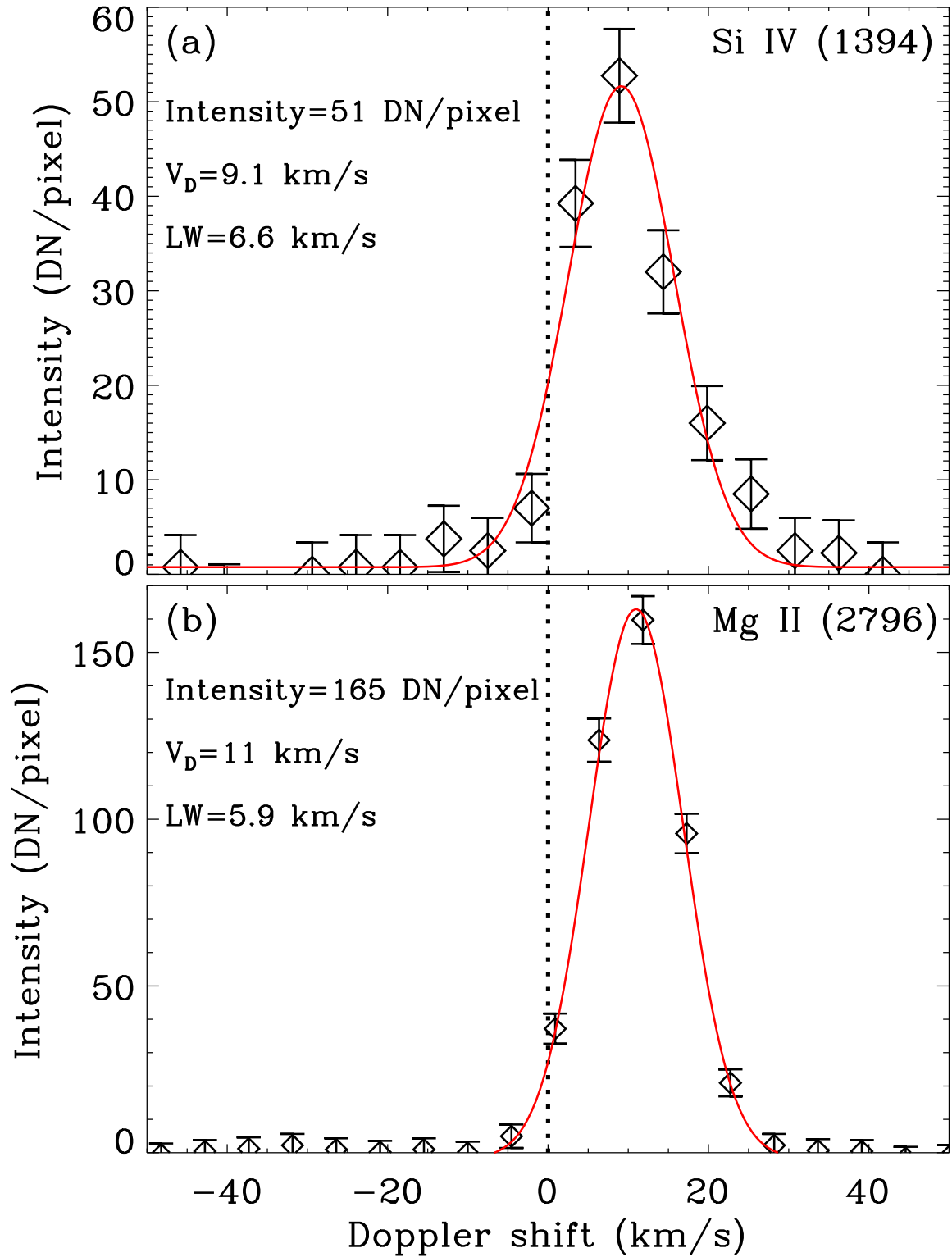}
\centering
\caption{Sample spectral profiles of the coronal rain observed by IRIS. (a) Si IV (1394 \AA) and (b) Mg II k (2796 \AA) line profiles of the coronal rain marked by the upper-left blue square in Figure \ref{f:cr_iris}(a). The diamonds (with the bars) show the observed spectra, and the red lines display the single-Gaussian fits to the spectra. The vertical dotted lines mark zero Doppler shift. The parameters denoted by the numbers in the plots show the peak intensity, Doppler shift (V$_{D}$), and Gaussian line width (LW) of the fits. See Section \ref{sec:cr_iris} for details.
\label{f:lp}}
\end{figure}

\begin{figure}[ht!]
\centering
\includegraphics[width=0.6\textwidth]{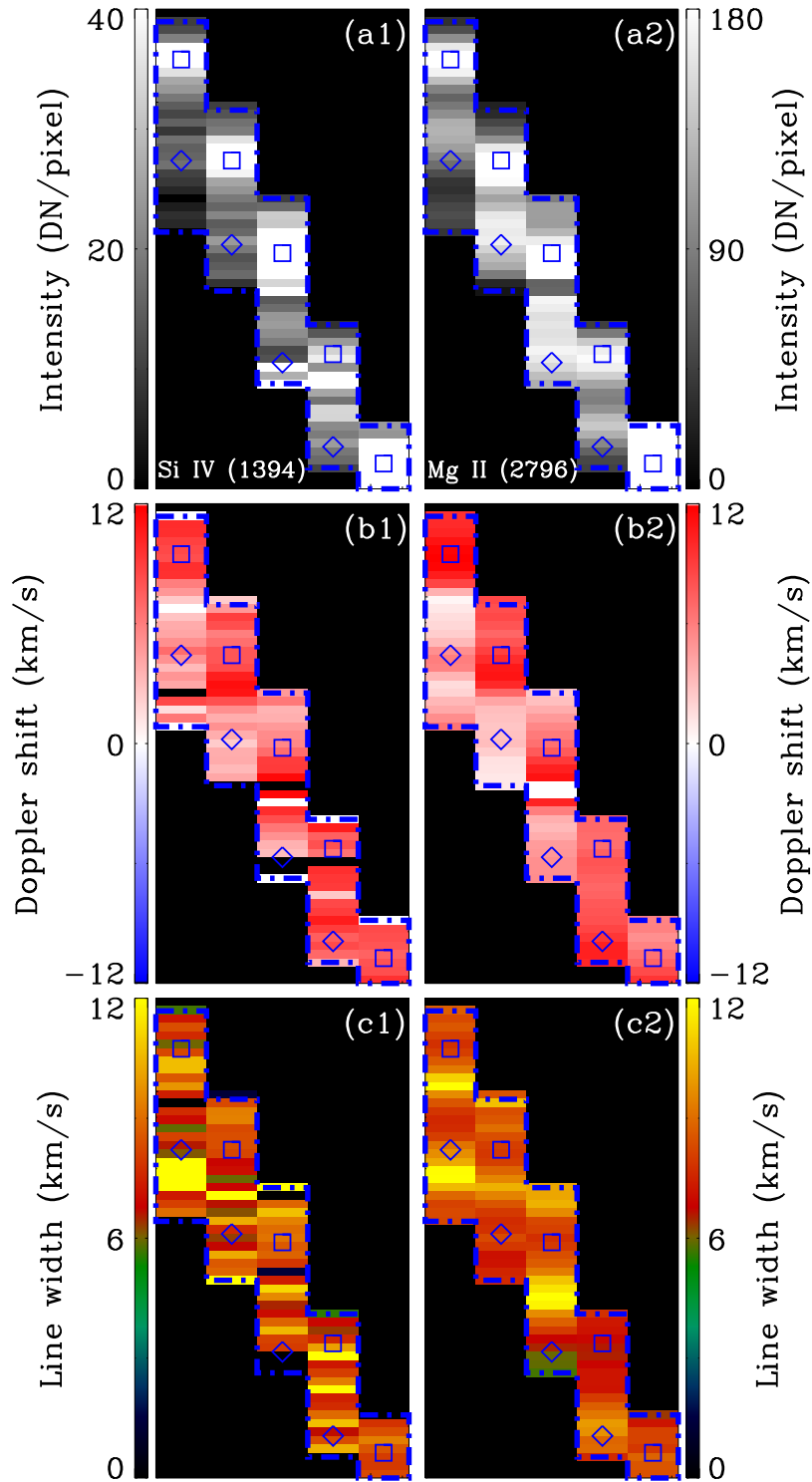}
\caption{Spectral information of the coronal rain observed by IRIS. Maps of intensity (a1)-(a2), Doppler shift (b1)-(b2), and line width (c1)-(c2) of Si IV (1394 \AA) (a1)-(c1) and Mg II k (2796 \AA) (a2)-(c2) lines for the IRIS raster scan \#3. Like in Figure \ref{f:cr_iris}(a), the blue rectangles and diamond separately indicate the positions of northern and southern parts of the coronal rain, which is enclosed by the blue dash-dotted lines. The FOV is denoted by the pink rectangle in Figure \ref{f:cr_iris}(a). See Section \ref{sec:cr_iris} for details.
\label{f:gf}}
\end{figure}

In the IRIS spectra, we see the coronal rain not only in Si IV, but also in the Mg II k line; see Section \ref{sec:cr_iris}. The coronal rain, therefore, also contains plasma at chromospheric temperatures of only $\sim$0.01 MK (Mg II k). This is in line with the observations of coronal rain in active region closed loops \citep{2015ApJ...806...81A} and suggests that the multithermal character of coronal rain is a common property to all types of coronal rains.  We find emission line profile of Mg II k, with no central reversal in the coronal rain (see Figure \ref{f:lp}(b)). This is suggestive of a reduced opacity \citep{2015ApJ...803...85L}, also supported by the line ratios of Mg II 2796 \AA~and 2803 \AA; see Section \ref{sec:cr_iris}. We find small but similar Doppler shifts of both the Si IV and Mg II k lines to the red, with values of 7.5-8.7 (5-5.9) km s$^{-1}$ for the north (south) coronal rain; see Section \ref{sec:cr_iris}. Considering a simple geometry, using these  Doppler shifts and the typical plane-of-sky (POS) speed of 40 km s$^{-1}$ (see Figure \ref{f:cr_iris}), we can calculate the inclination angle of the coronal rain, and thus the angle of the magnetic field channeling the downward flows.
We find 10.6$^{\circ}$-12.3$^{\circ}$ (7.1$^{\circ}$-8.4$^{\circ}$) behind the POS for the north (south) parts of the coronal rain. Thus, the plane of the loops seen here close to the limb is nearly vertical.
Consequently, the POS speeds of the coronal rain represent (almost) the actual speed. Moreover, the widths of both the Si IV and Mg II k lines are very small, close to the thermal widths and much smaller than those observed usually \citep{2015ApJ...799L..12D,2019A&A...626A..98L}; see Figures \ref{f:gf}(c1)-(c2). For Si IV in the coronal rain we typically find a line width of $\sim$8 km s$^{-1}$ with the thermal width being 6.86 km s$^{-1}$, smaller than those for quiescent coronal rain in the active region closed loops \citep{2014cosp...40E.105A} and along the open loops \citep{2015ApJ...807....7R}, indicating a reduced level of non-thermal motions.

We investigated the coronal context of the rain observed by IRIS through the AIA EUV images. Seven repeated condensations facilitated by repeated interchange reconnection between a system of open and closed loops are found during three days from 2013 October 18 to 20. Our results concerning the coronal evolution are similar to the findings of \citet{2018ApJ...864L...4L, 2018ApJ...868L..33L, 2019ApJ...884...34L}, in particular with respect to the motions and reconnection of loops, the formation of dips, the cooling and formation of condensations, and the downflows of condensations; see Section \ref{sec:cr_sdo}. 
In our earlier studies \citep{2018ApJ...864L...4L, 2018ApJ...868L..33L, 2019ApJ...884...34L} we could follow the cooling of the plasma only down to the characteristic temperatures captured by the 304\,{\AA} channel, i.e.,\ about 0.05\,MK. In our new study we add to this picture the evolution of the condensations cooling through the transition region (Si IV) all the way to chromospheric temperatures captured by the Mg II line. 
Our results further support the new and alternative formation mechanism for coronal rain due to reconnection along open field lines \citep{2018ApJ...864L...4L}. 

The seven condensations, and hence the coronal rain events, take place repeatedly in the same manner. We found the occurrence interval of coronal rain to be in the range of 4.6-10.4 hr with a mean value of $\sim$6.6 hr. This apparent quasi-periodicity is consistent with the periods of quiescent coronal rain in active region closed loops \citep{2012ApJ...745..152A, 2020A&A...633A..11F}. In contrast to those previous studies, our analysis underlines the fundamental role that reconnection plays in the mass cycle of coronal plasma by facilitating the recurring coronal rain. \citet{2019ApJ...874L..33M} suggest that the periodicity could be explained by a combination of thermal non-equilibrium (in the closed, underlying loops) and interchange reconnection with the open loops. As thermal non-equilibrium is not expected to occur on open field lines \citep{2019ApJ...874L..33M}, it cannot be used to explain the quasi-periodicity of coronal rain that occurs along the open loops in this study. Here, the global topological changes caused by reconnection clearly play a role, as the repetition observed in the coronal rain closely follows the formation of dips. 

If we would consider the IRIS observations alone, the coronal rain events we report here wound resemble those occurring in magnetically closed loops. Commonly, those have been interpreted in the framework of heating-condensation cycles due to the loss of thermal equilibrium \citep{2001SoPh..198..325S, 2003A&A...411..605M, 2005A&A...436.1067M, 2010ApJ...716..154A, 2020PPCF...62a4016A}. However, combining the observations of IRIS and AIA, we find clear evidence that the coronal rain in IRIS corresponds to the downflows of condensation, i.e., coronal rain, facilitated by interchange reconnection between loops in AIA; see Section \ref{sec:relation}. 
Therefore, the detection of coronal rain along loop-like structures cannot be uniquely attributed to thermal non-equilibrium and the quiescent coronal rain of the active region kind, unless no null-point topology is present. Otherwise, interchange reconnection and the presence of magnetic dips play an important role in coronal rain formation.

\clearpage
\startlongtable
\begin{deluxetable}{c c c c c c c c}
\tabletypesize{\scriptsize}
\tablecaption{General information on the IRIS raster scans and slit-jaw images (SJIs). 
\label{tab:iris}}
\tablehead{
\multicolumn{2}{c}{2013 October} & \multicolumn{6}{c}{Spectrograph Raster Scans}  \\
\cline{3-8}
Start Time & End Time & Center & FOV & Spatial Sampling & Step & Exposure & Steps \\
(UT) & (UT) & (\arcsec) & (\arcsec) & along Slit (\arcsec/pixel) & (s)  & Time (s)
}
\startdata
18 23:59:51 & 19 00:32:46 & -783.6, -621.8 & 126.8$\times$128.4  & 0.333 & 31.4 & 30 & 64$\times$2\arcsec \\ 
19 00:33:19 & 19 01:06:14 & -783.5, -622.0 & 126.8$\times$128.4 & 0.333 & 31.4 & 30 & 64$\times$2\arcsec \\
19 01:06:46 & 19 01:39:41 & -783.5, -622.1 & 126.8$\times$128.4 & 0.333 & 31.4 & 30 & 64$\times$2\arcsec \\
19 01:40:13 & 19 02:13:08 & -783.6, -621.5 & 126.8$\times$128.4 & 0.333 & 31.4 & 30 & 64$\times$2\arcsec \\
19 02:13:40 & 19 02:46:36 & -783.5, -621.9 & 126.8$\times$128.4 & 0.333 & 31.4 & 30 & 64$\times$2\arcsec \\
19 02:47:08 & 19 03:20:03 & -783.5, -621.9 & 126.8$\times$128.4 & 0.333 & 31.4 & 30 & 64$\times$2\arcsec \\
\hline
 \multicolumn{2}{c}{2013 October} &  \multicolumn{6}{c}{SJIs (1400 \AA)}  \\
 \cline{3-8}
Start Time & End Time & Center & FOV & Spatial Sampling & Time Cadence & Exposure \\
(UT) & (UT) & (\arcsec) & (\arcsec) & (\arcsec/pixel) & (s) & Time (s) \\
 \hline
18 23:59:51  & 19 03:20:03 & -781.3, -662.4 & 245.5$\times$128.4 & 0.333 & 31.4 & 30 \\
\enddata
\tablecomments{The 18 and 19 in the front of the columns 1-2 represent the dates of 2013 October 18 and 19, respectively.}
\end{deluxetable}

\startlongtable
\begin{deluxetable}{c c c c}
\tabletypesize{\scriptsize}
\tablecaption{Lines of interest. 
\label{tab:lines}}
\tablehead{
\colhead{Line} & \colhead{Rest Wavelength (\AA)} & \colhead{$\Delta$v (km s$^{-1}$)} & \colhead{Doppler Shift (km s$^{-1}$)} 
}
\startdata
S I & 1392.5878 & -251.243 & -0.4   \\ 
Fe II & 1392.817 & -201.898 & -0.9   \\ 
Ni II & 1393.33 & -91.49 & 1.3 \\
Si IV (1394) & 1393.755 & 0 & 2.3 \\
\hline
Mn I & 2795.64 & -75.8523 & 0.4   \\ 
Mg II (2796) & 2796.347 & 0 & 3.1   \\ 
Fe I & 2798.5999 & 241.701 & -0.3 \\
Ni I & 2799.476 & 335.704 & -0.0 \\
\enddata
\tablecomments{Listed are the lines, the rest wavelengths of the lines, the Doppler shifts ($\Delta$v) of the rest wavelengths of the lines to the Si IV (upper) and Mg II k (lower) lines, and the absolute Doppler shifts of the lines in the spectrum averaged over a quiet Sun region after the wavelength calibrations. Positive (negative) values of the Doppler shifts correspond to redshifts (blueshifts). The rest wavelengths of the lines are obtained from the line list of \citet{2017RAA....17..110T}.}
\end{deluxetable}

\startlongtable
\begin{deluxetable}{c c c c c c c c c}
\tabletypesize{\scriptsize}
\tablecaption{Measurements of coronal condensations from 2013 October 18 to 20. 
\label{tab:cc}}
\tablehead{
  &  &   & \multicolumn{2}{c}{Cooling Times}  &   &   &   \\
\cline{4-5}
  &  Start Time & End Time & 171-131 \AA & 131-304 \AA & Appearance & Peak Time & Disappearance & Lifetime \\
Event & (UT) & (UT) & (min) & (min) & Time (UT) & (UT) & Time (UT) & (hr)
}
\startdata
1 & 18 10:00 & 18 19:00 & 25 & 82 & 18 14:20 & 18 16:24 & 18 18:50 & 4.5 \\
2 & 18 19:00 & 18 23:45 & 30 & 48 & 18 19:10 & 18 21:57 & 18 23:40 & 4.5 \\
3 & 18 23:45   & 19 05:15 & 11  & 118 & 18 23:53 & 19 02:06 & 19 05:13 & 5.3 \\
4 & 19 05:15 & 19 13:10  & 42 & 81 & 19 05:23 & 19 07:29 & 19 13:10 & 7.8  \\
5 & 19 13:10  & 19 19:30  & 51  & 162 & 19 13:10 & 19 17:51 & 19 19:30 & 6.3 \\
6 & 19 19:30  & 20 02:40  & 24 & 105 & 19 19:30 & 19 23:06 & 20 02:40 & 7.2 \\
7 & 20 02:40  & 20 13:00  & 51 & 174 & 20 04:55 & 20 07:51 & 20 13:00 & 8.1 \\
\enddata
\tablecomments{Events 1-4 are displayed in Figures \ref{f:cc}(a1)-(c1), (a2)-(c2), (a3)-(c3), and (a4)-(c4), respectively. Same to Table \ref{tab:iris}, the 18, 19, and 20 in the front of the columns 2-3 and 6-8 separately represent the dates of 2013 October 18, 19, and 20.}
\end{deluxetable}

\clearpage

\clearpage

\clearpage

\clearpage

\clearpage

\clearpage

\clearpage

\clearpage

\clearpage

\acknowledgments
The authors would like to acknowledge valuable input from Dr. Patrick Antolin, who served as a reviewer. We are indebted to the SDO and IRIS teams for providing the data. AIA data are the courtesy of NASA/SDO and the AIA, EVE, and HMI science teams. IRIS is a NASA small explorer mission developed and operated by LMSAL with mission operations executed at NASA Ames Research center and major contributions to downlink communications funded by ESA and the Norwegian Space Centre. This work is supported by the B-type Strategic Priority Program of the Chinese Academy of Sciences  (XDB41000000), the National Natural Science Foundations of China (11533008, 11673034, 12073042, 11790304, 11873059, 1111903050, and 11773039), and the Key Research Program of Frontier Sciences (ZDBS-LY-SLH013) and the Key Programs (QYZDJ-SSW-SLH050) of the Chinese Academy of Sciences. This project is supported by the Specialized Research Fund for Shandong Provincial Key Laboratory. We acknowledge the usage of JHelioviewer software \cite[][]{2017A&A...606A..10M}. This research has made use of NASA's Astrophysics Data System. 


\end{document}